\documentclass[12pt]{article}
\usepackage{latexsym,amsmath}
\input amssym
\begin{document}
\renewcommand{\theequation}{\thesection.\arabic{equation}}
\def\prg#1{\medskip{\bf #1}}
\def\lra{\leftrightarrow}        \def\Ra{\Rightarrow}
\def\nin{\noindent}              \def\pd{\partial}
\def\dis{\displaystyle}          \def\dfrac{\dis\frac}
\def\grl{{GR$_\Lambda$}}         \def\vsm{\vspace{-9pt}}
\def\Lra{{\Leftrightarrow}}      \def\ads3{AdS$_3$}
\def\cs{{\scriptscriptstyle \rm CS}}  \def\ads3{{\rm AdS$_3$}}
\def\Leff{\hbox{$\mit\L_{\hspace{.6pt}\rm eff}\,$}}
\def\inn{\,\rfloor\,}
\def\bull{\raise.25ex\hbox{\vrule height.8ex width.8ex}}
\def\Tr{\hbox{\rm Tr\hspace{1pt}}}
\def\bF{{\bar F}}     \def\bt{{\bar\tau}}
\def\att#1{~{{\Large$\bullet$}~#1}}
\def\as{{\rm as}}

\def\D{{\Delta}}      \def\bC{{\bar C}}     \def\bT{{\bar T}}
\def\bH{{\bar H}}     \def\bL{{\bar L}}     \def\bB{{\bar B}}
\def\hO{{\hat O}}     \def\hG{{\hat G}}     \def\tG{{\tilde G}}
\def\cL{{\cal L}}     \def\cM{{\cal M }}    \def\cE{{\cal E}}
\def\cA{{\cal A}}     \def\cI{{\cal I}}     \def\cC{{\cal C}}
\def\cF{{\cal F}}     \def\hcF{\hat{\cF}}   \def\bcF{{\bar\cF}}
\def\cH{{\cal H}}     \def\hcH{\hat{\cH}}   \def\bcH{{\bar\cH}}
\def\cK{{\cal K}}     \def\hcK{\hat{\cK}}   \def\bcK{{\bar\cK}}
\def\cO{{\cal O}}     \def\hcO{\hat{\cal O}} \def\tR{{\tilde R}}
\def\cB{{\cal B}}     \def\bV{{\bar V}}

\def\G{\Gamma}        \def\S{\Sigma}        \def\L{{\mit\Lambda}}
\def\a{\alpha}        \def\b{\beta}         \def\g{\gamma}
\def\d{\delta}        \def\m{\mu}           \def\n{\nu}
\def\th{\theta}       \def\k{\kappa}        \def\l{\lambda}
\def\vphi{\varphi}    \def\ve{\varepsilon}  \def\p{\pi}
\def\r{\rho}          \def\Om{\Omega}       \def\om{\omega}
\def\s{\sigma}        \def\t{\tau}          \def\eps{\epsilon}
\def\ups{\upsilon}    \def\tom{{\tilde\om}} \def\bw{{\bar w}}
\def\nab{\nabla}      \def\tnab{{\tilde\nabla}}
\def\tcR{\tilde{\cal R}}
\def\Th{\Theta}       \def\cT{{\cal T}}    \def\cS{{\cal S}}
\def\nul{{\hat 0}}   \def\one{{\hat 1}}   \def\two{{\hat 2}}
\newcommand{\hodge}{{}^\star}
\def\nn{\nonumber}
\def\be{\begin{equation}}             \def\ee{\end{equation}}
\def\ba#1{\begin{array}{#1}}          \def\ea{\end{array}}
\def\bea{\begin{eqnarray} }           \def\eea{\end{eqnarray} }
\def\beann{\begin{eqnarray*} }        \def\eeann{\end{eqnarray*} }
\def\beal{\begin{eqalign}}            \def\eeal{\end{eqalign}}
\def\lab#1{\label{eq:#1}}             \def\eq#1{(\ref{eq:#1})}
\def\bsubeq{\begin{subequations}}     \def\esubeq{\end{subequations}}
\def\bitem{\begin{itemize}}           \def\eitem{\end{itemize}}

\title{Self-dual Maxwell field in 3D gravity with torsion}

\author{M. Blagojevi\'c and B. Cvetkovi\'c\footnote{
        Email addresses: {\tt mb@phy.bg.ac.yu,
                                cbranislav@phy.bg.ac.yu}} \\
Institute of Physics, P. O. Box 57, 11001 Belgrade, Serbia}
\date{\small\today, sd5.tex }
\maketitle
\begin{abstract}
We study the system of self-dual Maxwell field coupled to 3D gravity
with torsion, with Maxwell field modified by a topological mass term.
General structure of the field equations reveals a new, dynamical role
of the classical central charges, and gives a simple correspondence
between self-dual solutions with torsion and their Riemannian
counterparts. We construct two exact self-dual solutions, corresponding
to the sectors with massless and massive Maxwell field, and calculate
their conserved charges.
\end{abstract}

\section{Introduction}
\setcounter{equation}{0}

The three-dimensional (3D) gravity has been used for nearly three
decades as a laboratory for exploring basic features of the realistic
gravitational dynamics, with a number of outstanding results \cite{1}.
In particular, one should mention here the discovery of the Ba\~nados, Teitelboim and
Zanelli (BTZ) black
hole \cite{2}, the study of which helped us to improve our basic
understanding of the classical and quantum black hole dynamics.

In the early 1990s, Mielke and Baekler (MB) \cite{3} introduced a new
element into this structure by replacing the traditional Riemannian
geometry of general relativity (GR) by the Riemann-Cartan geometry,
in which the gravitational dynamics is characterized by both the
curvature and the torsion. Such an approach is expected to give us a
new insight into the relationship between geometry and the dynamical
structure of gravity. Recent developments along these lines led to a
number of interesting results related to the Chern-Simons
formulation, conformal asymptotic structure, black hole solutions,
thermodynamics and supersymmetry \cite{4,5,6,7}, which confirm that
the topological MB model has a reach dynamical structure. Here, we
would like to single out one of these results---that 3D gravity with
torsion possesses the BTZ-like black hole solution \cite{4}, which is
electrically neutral. In an attempt to extend this result to the
electrically charged sector, we analyzed {\it static\/}
configurations of the MB model and found an exact solution with
azimuthal electric field \cite{8}. In the present paper, we continue
this investigation of the charged sector by constructing two {\it
rotating\/} solutions, corresponding to the {\it self-dual\/} Maxwell
field as a source. Standard dynamics of the Maxwell field is modified
by introducing a topological, gauge-invariant mass term \cite{9},
which has a ``screening" effect and eliminates logarithmic
divergences. The results obtained here can be compared with similar
investigations in \grl\ \cite{10,11,12,13,14}, with a possibility to
recognize dynamical effects of torsion.

The paper is organized as follows. In section 2, we give a brief
account of 3D gravity with torsion and examine general structure of
the field equations, with Maxwell field modified by a topological
mass  term ($\m$). In section 3, we restrict the Maxwell field to be
self-dual and examine the related dynamical structure. As a result,
we discovere a significant dynamical role of the classical central
charge of 3D gravity with torsion \cite{5}: two possible values of
the central charge are directly related to the two self-duality
sectors of the Maxwell field. Moreover, we prove a general theorem
which establishes a simple correspondence between self-dual solutions
with torsion, and their Riemannian counterparts. In sections 4 and 5,
relying on the results of Appendix A, we construct two exact,
stationary and spherically symmetric solutions, characterized by
$\m=0$ and $\m\ne 0$. They are recognized as natural generalizations
of the Kamata-Koikawa \cite{10} and Fernando-Mansouri \cite{13}
solutions, respectively, found earlier in Riemannian theory. For both
of these solutions, we use the canonical approach to calculate their
conserved charges---energy, angular momentum and electric charge.
Section 5 is devoted to concluding remarks. In Appendix A, we show
that every self-dual solution is determined by a single function of
radial coordinate (compare with \cite{14}), while Appendix B contains
some technical details.

Our conventions are given by the following rules: the Latin indices
refer to the local Lorentz frame, the Greek indices refer to the
coordinate frame;  the middle alphabet letters
$(i,j,k,...;\m,\n,\l,...)$ run over 0,1,2, the first letters of the
Greek alphabet $(\a,\b,\g,...)$ run over 1,2; the metric components in
the local Lorentz frame are $\eta_{ij}=(+,-,-)$; totally antisymmetric
tensor $\ve^{ijk}$ and the related tensor density $\ve^{\m\n\r}$ are
both normalized so that $\ve^{012}=1$; the Hodge-star operation is
$\hodge$.

\section{The field equations} 
\setcounter{equation}{0}

We begin this section with a brief account of the basic structure of
3D gravity with torsion, then we introduce the action containing the
Maxwell field modified by a topological (Chern-Simons) mass term and
derive the general form of the field equations.

Theory of gravity with torsion can be naturally described as a
Poincar\'e gauge theory (PGT), with an underlying spacetime structure
corresponding to Riemann-Cartan geometry. Basic gravitational
variables in PGT are the triad $b^i$ and the Lorentz connection
$A^{ij}=-A^{ji}$ (1-forms), and the corresponding field strengths are
the torsion $T^i$ and the curvature $R^{ij}$. In 3D, one can
introduce the notation $A^{ij}=:-\ve^{ij}{_k}\om^k$ and
$R^{ij}=:-\ve^{ij}{_k}R^k$, which yields:
\be
T^i=db^i+\ve^i{}_{jk}\om^j\wedge b^k \, ,\qquad
R^i=d\om^i+\frac{1}{2}\,\ve^i{}_{jk}\om^j\wedge\om^k\, .   \lab{2.1}
\ee

PGT is characterized by a useful identity:
\bsubeq\lab{2.2}
\be
\om^i\equiv\tom^i+K^i\, ,                                  \lab{2.2a}
\ee
where $\tom^i$ is the Levi-Civita (Riemannian) connection, and $K^i$
is the contortion 1-form, defined implicitly by
$T^i=:\ve^i{}_{mn}K^m\wedge b^n$. Using this identity, one can
express the curvature $R_i=R_i(\om)$ in terms of its {\it
Riemannian\/} piece $\tR_i=R_i(\tom)$ and the contortion $K_i$:
\be
2R_i\equiv 2\tR_i+2\tnab K_i+\ve_{imn}K^m\wedge K^n\, .    \lab{2.2b}
\ee
\esubeq
The covariant derivative $\nabla=\nabla(\om)$ acts on a tangent-frame
spinor/tensor in accordance with its spinorial/tensorial structure;
when $X$ is a form, $\nabla X:=\nabla\wedge X$, and
$\tnab=\nab(\tom)$.

The antisymmetry of the Lorentz connection $A^{ij}$ implies that the
geometric structure of PGT corresponds to Riemann-Cartan geometry, in
which $b^i$ is an orthonormal coframe, $g:=\eta_{ij}b^i\otimes b^j$
is the metric of spacetime, $\om^i$ is the Cartan connection, and
$T^i,R^i$ are the torsion and the Cartan curvature, respectively. In
what follows, we will omit the wedge product sign $\wedge$ for
simplicity.

General gravitational dynamics in Riemann-Cartan spacetime is
determined by Lagrangians which are at most quadratic in field
strengths. Omitting the quadratic terms, we arrive at the {\it
topological\/} MB model for 3D gravity \cite{3}:
\bsubeq\lab{2.3}
\be
L_0=2ab^i R_i-\frac{\L}{3}\,\ve_{ijk}b^i b^j b^k\
    +\a_3L_\cs(\om)+\a_4 b^i T_i\, ,                       \lab{2.3a}
\ee
where $a=1/16\pi G$ and $L_\cs(\om)=\om^id\om_i
+\frac{1}{3}\ve_{ijk}\om^i\om^j\om^k$ is the Chern-Simons Lagrangian
for the Lorentz connection. The MB model is a natural generalization
of GR with a cosmological constant (\grl). The complete dynamics of
the topologically massive Maxwell field coupled to 3D gravity with
torsion is described by the Lagrangian
\be
L=L_0+L_M\, ,\qquad
L_M:=-\frac{1}{2}F{\,}\hodge F-\frac{\mu}{2}AF\, ,         \lab{2.3b}
\ee
\esubeq
where $F=dA$.

By varying $L$ with respect to $b^i$ and $\om^i$, one obtains the
gravitational field equations. In the nondegenerate sector with
$\D:=\a_3\a_4-a^2\neq 0$, they have the form \cite{8}
\bsubeq\lab{2.4} \bea
&&2T_i-p\ve_{ijk}b^jb^k=u\Th_i\, ,                         \lab{2.4a}\\
&&2R_i-q\ve_{ijk}b^jb^k=-v\Th_i\, ,                        \lab{2.4b}
\eea
\esubeq
where $\Th_i=-\d L_M/\d b^i$ is the Maxwell energy-momentum current
(2-form), and
\bea
&&p:=\frac{\a_3\L+\a_4 a}{\D}\,,\qquad u:=\frac{\a_3}{\D}\,,\nn\\
&&q:=-\frac{(\a_4)^2+a\L}{\D}\,,\qquad  v:=\frac{a}{\D}\,. \nn
\eea
After introducing the the energy-momentum tensor,
$$
\cT^i{_k}:=\hodge(b^i\Th_k)=-F^{im}F_{km}+\frac{1}{4}\d^i_k F^2\, ,
$$
where $F^2=F^{mn}F_{mn}$, we can express $\Th_i$ as
$$
\Th_i=\ve_{imn}t^mb^n\, ,\qquad
   t^m:=-\left(\cT^m{_k}-\frac{1}{2}\d^m_k\cT\right)b^k\, ,
$$
with $\cT=\cT^k{_k}$, whereupon the gravitational field equations
take the simple form:
\bsubeq\lab{2.5}
\bea
&&T_i=\ve_{imn}K^mb^n\, ,\qquad
  K^m=\frac{1}{2}(pb^m+ut^m)\, ,                           \lab{2.5a}\\
&&2R_i=q\ve_{imn}b^mb^n-v\ve_{imn}t^mb^n\, .               \lab{2.5b}
\eea
These equations, together with the modified Maxwell equations
\be
d{\,}\hodge F+\m F=0\, ,                                   \lab{2.5c}
\ee
\esubeq
define the complete dynamics of our system. The Cartan curvature $R_i$
is calculated using the identity \eq{2.2b},
\be
2R_i=2\tR_i+u\tnab t_i
 +\ve_{imn}\left(\frac{p^2}{4}b^mb^n+\frac{up}{2}t^mb^n
 +\frac{u^2}{4}t^mt^n\right)\, ,                           \lab{2.6}
\ee
where the Maxwell field contribution is compactly represented by the
1-form $t^i$.

\section{Dynamical characteristics of self-dual solutions}
\setcounter{equation}{0}

General structure of the field equations \eq{2.5} depends essentially
on the form of the Maxwell field. In this section, we discuss some
important dynamical characteristics of exact solutions corresponding
to the self-dual Maxwell field as a source.

We are looking for a {\it spherically symmetric\/} and {\it
stationary\/} solution. Choosing the local coordinates
$x^\m=(t,r,\vphi)$, we make the following ansatz for the triad
field,
\be b^0=Ndt\,,\qquad b^1=B^{-1}dr\,,\qquad
b^2=K(d\vphi+Cdt)\,,\lab{3.1} \ee and for the Maxwell field: \be
F=Eb^0b^1-Hb^1b^2\, .
\lab{3.2} \ee Here, $N,B,C,K$ and $E,H$ are six unknown functions
of the radial coordinate $r$.

General form of the modified Maxwell field equations \eq{2.5c} reads:
\bea
&&E'B+\g E+\m H=0\, ,                                      \nn\\
&&H'B+\a H+2\b E+\m E=0\, ,                                \lab{3.3}
\eea
where prime denotes the derivative with respect to the radial
coordinate $r$, and $\a,\b,\g$ are components of the Riemannian
connection, defined in \eq{A1}. We assume a generalized {\it
self-duality\/} of the Maxwell field:
\be
E=\eps H\, , \qquad \eps^2=1\, .                           \lab{3.4}
\ee
Taking the difference of the two Maxwell equations in conjunction
with the self-duality condition leads to
\bsubeq\lab{3.5}
\be
2\b\eps=\g-\a\, .
\ee
Introducing a new radial coordinate $\r=\r(r)$, defined by
$$
d\r=\frac{dr}{B}\, ,
$$
we find that the first integrals of these equations are given as
\bea
&&EK=-Q_e\exp(-\eps\m\r)\, ,                               \nn\\
&&-HN=(Q_m-Q_eC)\exp(-\eps\m\r)\, ,                        \lab{3.5b}
\eea
where $Q_e,Q_m$ are integration constants. Inserting here the
self-duality condition yields
\be
\eps Q_eN=K(Q_m-Q_eC)\, .
\ee
\esubeq

The self-duality condition implies $F^2=0$ and simplifies the form of
the electromagnetic energy-momentum tensor,
$$
\cT^i{_j}=\left(\ba{ccc}
            E^2 & 0 & EH \\
             0  & 0 & 0  \\
            -EH & 0 & -H^2
                 \ea \right) \, ,
$$
and the form of $t^i$:
\bea
&&t^i=-\cT^i{_j}b^j\, ,                                    \nn\\
&&t^0=-E(Eb^0+Hb^2)\, ,\qquad t^1=0\, ,\qquad t^2=-\eps t^0\, .\nn
\eea
This defines the contortion as in \eq{2.5a}, and exhausts the content
of the first field equation.

Going now to the second field equation \eq{2.5b}, we combine the
relations
$$
\ve_{imn}t^mt^n=0\, ,\qquad
\tnab t_i =\eps\frac{B(NK)'}{NK}\ve_{imn}t^mb^n\, ,
$$
with \eq{A5} and calculate the Cartan curvature \eq{2.6}:
\be
2R_i=2\tR_i+\frac{p^2}{4}\ve_{imn}b^mb^n
     +\left(\frac{pu}{2}+\eps\frac{u}{\ell}\right)\ve_{imn}t^mb^n\, .
\ee
Then, by substituting this result into \eq{2.5b}, we obtain:
\bea
&&2\tR_i=\Leff\ve_{imn}b^mb^n-V\ve_{imn}t^mb^n\, ,         \lab{3.7}\\
&&V:=v+\frac{pu}{2}+\eps\frac{u}{\ell}\, .                 \nn
\eea
Note that the factor $V=V(\eps)$ is proportional to the {\it
classical central charge\/} $c(\eps)$, characterizing the asymptotic
conformal structure of 3D gravity with torsion \cite{5}:
\bea
&&V(\eps)=\frac{1}{\D}\,\frac{1}{24\pi\ell}\,c(\eps)\, ,   \nn\\
&&c(\mp 1)=24\pi\left[a\ell
  + \a_3\left(\frac{p\ell}{2}\mp 1\right)\right]\, .       \nn
\eea
\bitem
\item[(a)] Equation \eq{3.7} reveals a new feature of the central
charges $c(\mp 1)$, showing that they have a direct dynamical
influence on the self-duality modes $\eps=\mp 1$ of the system.
\eitem

The set of equations \eq{3.7} consists of 9 ordinary differential
equations for only 3 unknown functions of $r$: $N,B$ and $K$ ($C$,
$E$ and $H$ are determined in terms of $N$ and $K$ as in \eq{3.5b}).
Is there any consistent solution of this overdetermined set of
equations? To answer this question, one can solve the system \eq{3.7}
directly, as shown in Appendix A, but we follow here another
approach, based on some simple properties of \eq{3.7}. Namely, in the
limit $u\to 0$, or equivalently, $V\to v=-1/a$, equation \eq{3.7}
becomes equivalent to the Einstein equation for the self-dual Maxwel
field in {\it Riemannian\/} 3D gra\-vi\-ty. This property can be
formulated as the following constructive statement:
\bitem
\item[(b)] Starting with any self-dual solution in Riemannian 3D
gravity, one can generate the related self-dual solution with torsion
by making the replacement $v\to V$.
\eitem
In what follows, we shall illustrate the power of this theorem by
constructing two different solutions of \eq{3.7}, belonging to the
sectors with $\m=0$ and $\m\ne 0$, respectively.

\section{A self-dual solution with \boldmath{$\m=0$}} 
\setcounter{equation}{0}

\subsection{Construction} 

The condition $\m=0$ refers to the standard, massless Maxwell field.
As shown in Appendix A, the field equations define the form of $K^2$
as in \eq{A6},
\bea
&&K^2=g_1+g_2\exp{(2\r/\ell)}+\frac{2}{\ell}r_0^2\r\, ,    \nn\\
&&r_0^2:=-\frac{\ell^2Q_e^2}{4}\,V(\eps)\, ,               \nn
\eea
whereupon the general solution follows from equations \eq{A4} and
\eq{3.5b}. To realize this construction, we first conveniently fix
$g_2$ by demanding $g_2=r_0^2$. Then, choosing the radial coordinate
$r$ by $B=(r^2-r_0^2)/\ell r$, we find
$$
\r=\int \frac{dr}{B}
  =\frac{\ell}{2}\ln\left|\frac{r^2-r_0^2}{r_0^2}\right|\, .
$$
Finally, using \eq{A4} and \eq{3.5b}, we obtain the general solution
for $\m=0$:
\bsubeq\lab{4.1}
\bea
&&B=\frac{r^2-r_0^2}{\ell r}\, ,\qquad
  N=\frac{r^2-r_0^2}{\ell K}\, ,                           \nn\\
&&K^2=r^2+r_0^2\ln\left|\frac{r^2-r_0^2}{r_0^2}\right|+h_1\,,\nn\\
&&C=\frac{Q_m}{Q_e}-\eps\frac{N}{K}\, ,
  \qquad E=\eps H=-\frac{Q_e}{K}\, ,                       \lab{4.1a}
\eea
where $h_1=g_1-r_0^2$. The boundary condition $C\to 0$ for
$r\to\infty$ yields $Q_m/Q_e=1/\ell$, and consequently:
\be
C=\frac{\eps}{\ell}\left(1-\frac{r^2-r_0^2}{K^2}\right) \, .
\ee
To complete the solution, we display also the electromagnetic
potential:
\be
A=\left(\frac{Q_e}{2}\ln\left|\frac{r^2-r_0^2}{r_0^2}\right|
  +h_2\right)(dt+\eps\ell d\vphi)\, .                      \lab{4.1c}
\ee
\esubeq

In the limit $u\to 0$, or $V\to-1/a$, the solution \eq{4.1} coincides
with the Kamata-Koikawa self-dual solution, found in Riemannian \grl\
\cite{10} (see also \cite{12}). Thus, the field configuration
\eq{4.1} represent a generalization of the Kamata-Koikawa solution to
the self-dual solution with torsion, in accordance with the theorem
(b), Section 3. Our result also confirms that Kamata and Koikawa
indeed found a correct solution, in contrast to the opinion presented
in \cite{11}. In the limit $Q_e\to 0$, the solution \eq{4.1} reduces
asymptotically to the vacuum state of the BTZ-like black hole with
torsion \cite{5}.

The torsion and the Cartan curvature of the self-dual solution
\eq{4.1} can be calculated with the help of equations \eq{2.5}. It
follows that the scalar Cartan curvature is $R=-6q$, while
$\tR=-6\Leff$. Although $\tR$ is constant, the form of $\tR_i$
implies that the metric of the solution is not maximally symmetric.
The logarithmic function appearing in the solution stems from the
dimensionality of spacetime.

\subsection{The conserved charges} 

To gain a deeper insight into the nature of the self-dual solution
\eq{4.1}, we now turn our attention to its conserved charges.

\prg{(A)} The family of solutions \eq{4.1} is parametrized by
$(Q_e,h_1,h_2)$. Considering the neutral limit $Q_e\to 0$ in
Riemannian \grl, Kamata and Koikawa \cite{10} concluded that one
should fix the parameter $h_1$ to zero, as it leads to $K^2\to r^2$
in this limit. Accepting the choice $h_1=0$, Chan \cite{10} used the
quasi-local formalism to calculate its energy and angular momentum.
Quasi-local charges are seen to have {\it logarithmic divergences\/}
for large $r$, stemming from the logarithmic behavior of the
electromagnetic potential in 3D. Our calculations confirmed this
result in the canonical formalism, where the logarithmic terms
produce {\it divergent surface terms\/} in the improved canonical
generator.

\prg{(B)} An interesting attempt to handle these divergences by a
suitable {\it regularization\/} procedure was proposed in \cite{15}.
In this procedure, we enclose the system in a circle $C_\as$ having a
large, but finite radius $r_\as$. This circle represents a
regularized spatial boundary at infinity, and the asymptotic region
is defined by $r\to r_\as$. Then, we make a choice of the boundary
conditions by fixing the values of dynamical variables at $C_\as$.
Finally, at the end of our calculations, we take the limit
$r_\as\to\infty$.

To be more specific, let us consider the regularization of $K$. After
introducing $C_\as$, we fix the integration constant $h_1$ to the
value $\bar h_1:=r_0^2\ln[r_0^2/(r_\as^2-r_0^2)]$, so that
\bsubeq\lab{4.2}
\be
K^2=r^2+r_0^2\ln\left|\frac{r^2-r_0^2}{r_\as^2-r_0^2}\right|\, .
\ee
This choice is equivalent to the boundary condition $K^2=r_\as^2$ at
$C_\as$. In other words, the logarithmic term is effectively eliminated
from the boundary $C_\as$. In a similar manner, we regularize the
Maxwell potential $A$ by choosing $h_2=\bar h_2:=(Q_e/2r_0^2)\bar h_1$:
\be
A=\frac{Q_e}{2}\ln\left|\frac{r^2-r_0^2}{r_\as^2-r_0^2}\right|
                 (dt+\eps\ell d\vphi)\, .
\ee
\esubeq
This is equivalent to the boundary condition $A=0$ at $C_\as$.

Going to the canonical formalism, we formulate the asymptotic
conditions as follows:
\bitem
\item[(i)] the fields  $b^i$, $\om^i$ and $A$ belong to the family
of self-dual configurations \eq{4.1}, parame\-trized by
$Q_e$, with $h_1=\bar h_1$ and $h_2=\bar h_2$,\vsm
\item [(ii)] asymptotic symmetries have well-defined canonical
generators.
\eitem
The meaning of these conditions is quite clear: (i) means that our
calculations refer to the regularized solution, while (ii) ensures
that we have a well-defined phase space.

As shown in Appendix B, using the asymptotic form of the regularized
solution, one can find the related asymptotic parameters, which
describe rigid time translations, axial rotations and $U(1)$
transformations. After that, the standard canonical procedure yields
the following expressions for the energy $E$, the angular momentum
$M$ and the electric charge $Q$ of our self-dual solution:
\bea
&&\ell E=\eps Q_e^2\frac{u}{12\ell}\,c(-\eps)=\eps M\, ,   \nn\\
&&Q=2\pi Q_e\, .                                           \lab{4.3}
\eea

Although the regularized self-dual solution has the same {\it
leading\/} asymptotic terms as the black hole with torsion
(compare Appendix B with \cite{5}), its conserved charges $E$ and
$M$ are quite different. The reason for this lies in the fact that
$E$ and $M$ are basically determined by the {\it sub-leading\/}
asymptotic terms (as noted by Chan \cite{10}). This is clearly
seen in the canonical formalism, where the analysis of the
relevant surface terms shows that $E$ and $M$ in \eq{4.3} stem
entirely from $ut^i$, the Maxwell field contribution to the
\emph{contortion\/}. Thus, the nonvanishing $E$ and $M$ are
generated by a combined effect of the Maxwell field ($t^i$) and
the gravitational Chern-Simons term ($u\sim\a_3$). This should be
compared with Riemannian \grl, where the regularization procedure
would yield $E=M=0$.

The regularized solution and the one with $h_1=h_2=0$, are two
different members of the same family \eq{4.1}, which have {\it
different conserved charges\/}. Another difference is found in their
geometric properties. Namely, the form of the Maxwell field in
\eq{4.1a} implies that $K$ must be real, and consequently, $K^2>0$. By
inspecting the regularized $K^2$, one finds that $K^2$ is positive for
$r=r_\as$, and it  has a zero at some $r=r_*<r_\as$, such that
$$
(r_*^2-r_0^2)\exp(r_*^2/r_0^2)=r_\as^2-r_0^2\, .
$$
We see that in the limit $r_\as\to\infty$, $r_*$ also goes to
infinity. Hence, the form the regularized solution \eq{4.2} is valid
only in the asymptotic region. One expects that a reasonable
extension of this region can be found by going to more suitable local
coordinates.

\section{A self-dual solution with \boldmath $\mu\neq 0$}
\setcounter{equation}{0}

\subsection{Construction}

The topological mass $\m$ is introduced to regularize the asymptotic
behavior of the Maxwell field. As shown in Appendix A, the solution
for $K^2$ in the case $\m\ne 0$ (with $\eps\mu\ell\neq -1$), takes
the form
\bsubeq\lab{5.1}
\be
K^2=g_1+g_2\exp(2\r/\ell)
    -\frac{r_0^2}{\mu^2\ell^2+\eps\mu\ell}\exp(-2\eps\m\r)\,,
\ee
where we used $r_0^2=-\ell^2 Q_e^2V/4$, as in the previous section.
This form of $K^2$, combined with \eq{A4} and \eq{3.5b}, leads to the
general self-dual solution with torsion (for $\m\ne 0$,
$\eps\m\ell\ne -1$):
\bea
&&N=\frac{g_3\exp(2\r/\ell)}{K}\, ,\qquad
  C=\frac{Q_m}{Q_e}-\eps\frac NK\, ,                       \nn\\
&&E=\eps H=-\frac{Q_e}{K}\exp(-\eps\m\r)\, ,               \lab{5.1b}
\eea
\esubeq
where $g_1$, $g_2$ and $g_3$ are constants of integration.

For a fixed $\eps$, the solution is characterized by eight
parameters: $g_1,g_2,g_3,\m,\ell,r_0,Q_e$ and $Q_m$. This number can
be significantly reduced by imposing various physical/geometric
requirements. To begin with, we normalize the time coordinate by
setting $g_3=\ell$. Next, demanding that $C$ vanishes at spatial
infinity, we obtain $Q_m/Q_e=\eps/\ell$. Then, we choose $g_2=\ell^2$
to ensure that in the limit $Q_e\to 0$, $g_1\to 0$, the solution
\eq{5.1} reduces to the black hole vacuum. Finally, we note that
solutions with vanishing Maxwell field at spatial infinity are
characterized by
\be
\eps\mu>0 \, .
\ee
As we shall see, this restriction ensures that we have finite
conserved charges (compare with \cite{14}). Without loosing
generality, we impose the requirement $\eps\mu= 1/\ell$, which
simplifies the calculations \cite{13} (see the comment at the end of
this section).

Since the solution \eq{5.1} is defined only up to a choice of the
radial coordinate, we impose the requirement $K=r$, which is
equivalent to
$$
\exp(2\r/\ell)=\frac{1}{2\ell^2}
   \left(r^2-g_1+\sqrt{(r^2-g_1)^2+2\ell^2 r_0^2}\right)\, .
$$
Expressed in terms of $r$, the solution \eq{5.1} reads:
\bea
&&K=r\, ,\qquad
  B=\frac{1}{\ell r}\sqrt{(r^2-g_1)^2+2\ell^2r_0^2}\, ,    \nn\\
&&N=\frac{1}{2\ell r}\left(r^2-g_1+\ell rB\right)\, , \qquad
  C=\eps\left(\frac1\ell-\frac Nr\right)\, ,               \nn\\
&&E=-\frac{Q_e}{r}\sqrt{\frac{\ell}{Nr}}\, ,\qquad
  A=-Q_e\sqrt{\frac\ell{Nr}}\left(dt+\eps\ell d\vphi\right)\,.\lab{5.3}
\eea
This result represents a natural generalization of the self-dual
solution found by Fernando and Mansouri \cite{13}, in the context of
Riemannian theory. An equivalent form of the Fernando-Mansouri
solution has been derived independently by Clem\'ent \cite{12}. In
the limit $Q_e\to 0$, the solution reduces to the extreme black hole
with torsion \cite{5} (see the next subsection). The parameter $g_1$
will be interpreted in terms of the conserved charges.

From the above expressions, we conclude that $B>0$ and $N>0$ over the
whole range of $r$, and moreover, the Cartan curvature is constant,
$R=-6q$. Hence, \eq{5.3} is a perfectly regular solution.

\subsection{The conserved charges}

As usual, our calculation of the conserved charges begins by fixing
the asymptotic conditions of the fields in \eq{5.3}:
\bsubeq\lab{5.4}
\bea
&&N\sim\frac{r}{\ell}-\frac{g_1}{\ell r}\, ,\qquad
  B\sim \frac{r}{\ell}-\frac{g_1}{\ell r}\, ,\qquad
  C\sim \frac{\eps g_1}{\ell r^2}\, ,                      \nn\\[3pt]
&&E\sim\frac{Q_e}{r^2}\,,\qquad A_\m\sim\frac{Q_e}{r}\, .  \lab{5.4a}
\eea

Note that the electromagnetic potential has the same asymptotic
behavior as in 4D.  As one can see from the absence of the
logarithmic terms, the topological mass $\m$ acts as an infrared
regulator, which modifies the asymptotics of all the fields.

Since $t^i\sim \cO_3$, one immediately concludes that the Maxwell
field contribution to the asymptotic behavior of $(b^i,\om^i)$ can be
ignored. In other words, due to the fast asymptotic decrease of the
Maxwell potential, the electromagnetic contribution to the energy and
angular momentum vanishes. Moreover, the asymptotics defined by
\eq{5.4a} coincides with that of the extreme BTZ black hole with
torsion \cite{5}, provided we make the identifications
\be
4G\ell m=\frac{g_1}{\ell}=4G\eps J\, .                     \lab{5.4b}
\ee
\esubeq
Using \eq{5.4b} and the expressions for the conserved charges of the
BTZ black hole with torsion \cite{5}, we find the energy and angular
momentum of our self-dual solution \eq{5.3}:
\bsubeq\lab{5.5}
\be
\ell E=\frac{g_1}{6\ell}\,c(-\eps)=\eps M\, .
\ee
In the Riemannian limit $u\to 0$, these expressions agree with those
obtained in the quasilocal formalism by Fernando and Mansouri
\cite{13}, and by Dereli and Obukhov \cite{14}.

To calculate the electric charge, we note that the canonical
generator of the theory with the topological mass $\m\ne 0$ contains
an additional term in the $U(1)$ sector. Namely, the $U(1)$ piece of
the canonical generator has the form (compare with Appendix A)
$$
G_3=\dot\l\pi^0-\l(\pd_\a\pi^\a-2\m\ve^{0\a\b}\pd_\a A_\b)\, ,
$$
where $\pi^\a=-bF^{0\a}-\mu\ve^{0\a\b}A_\b$. The variation of $G_3$
reads:
$$
\d G_3[\l]=2\pi\d\left(4Q_e\sqrt{\frac\ell{rN}}\right)\sim\cO_1\, ,
$$
hence $G_3$ is perfectly regular, and we have
\be
Q=0\, .
\ee
\esubeq
Thus, although the radial electric field exists, its asymptotic fall
off, given by $E\sim Q_e/r^2$, is too fast to produce a nonvanishing
electric charge. Consequently, the constant $Q_e$  can not be
interpreted as the electric charge, contrary to the expectation
expressed in \cite{13}. Thus, our Fernando-Mansouri-like solution
should be called a {\it neutral\/} self-dual solution.

\prg{Comment.} As we have seen, the assumption $\eps\m\ell=1$ allows
us to find a simple solution $\r=\r(r)$ of the condition $K=r$. In
the general case $\eps\m\ell\ne 1$, $\r(r)$ is not an elementary
functions. However, in the asymptotic region we have
$$
\exp\left(2\rho/\ell\right)\sim \frac{r^2-g_1}{\ell^2}\, ,
$$
so that
\be
B=\frac{dr}{d\rho}\sim\frac{r}\ell-\frac{g_1}{\ell r}\,.   \nn
\ee
Thus, the asymptotic form of $N$, $B$ and $C$ is the same as in the
case $\eps\m\ell=1$. Moreover, the asymptotic behavior of the Maxwell
variables reads:
\be
E\sim\frac{1}{r^{1+\eps\mu\ell}}\, ,\qquad
A_\mu\sim\frac{1}{r^{\eps\mu\ell}}\, .                     \nn
\ee
As a consequence, the conserved charges have again the values
\eq{5.5}. In other words, all solutions in the sector $\eps\mu>0$
have the same values of the conserved charges, as in Riemannian
theory \cite{14}.

\section{Concluding remarks}

In this paper, we studied dynamical properties and exact solutions of
the self-dual Maxwell field modified by a topological mass term, in
interaction with 3D gravity with torsion.

(1) General structure of the field equations implies the following
dynamical properties:
\bitem\vsm
\item[--] dynamical evolution is directly influenced by the classical
central charges, \vsm
\item[--] there is a simple correspondence between self-dual
solutions with torsion and their Riemannian counterparts, and \vsm
\item[--] any self-dual solution with torsion is completely determined
by a single function $K^2$.
\eitem\vsm

(2) We constructed two exact, stationary and self-dual solutions with
torsion, characterized by $\m=0$ and $\m\ne 0$. They represent
respective generalizations of the Kamata-Koikawa and
Fernando-Mansouri solutions, found earlier in Riemannian \grl. For
each of these solutions, we calculated its conserved charges. The
expressions for energy and angular momentum are proportional to the
classical central charges.

\appendix
\section*{Acknowledgements} 

This work was supported by the Serbian Science Foundation.

\section{Solving the gravitational field equations}
\setcounter{equation}{0}

In this Appendix, we demonstrate that the set of the gravitational
equations \eq{3.7} and the self-dual Maxwell equations \eq{3.3}
reduces to a simple differential equation for $K^2$. Integrating this
equation, we find that its general solution depends on $\m$.

The Riemannian connection $\tom^i$ is determined by the condition of
vanishing torsion, $db^i+\ve_{ijk}\tom^jb^k=0$. Starting from the
expression \eq{3.1} for $b^i$, we find:
\bea
&&\tom^0=-\b b^0-\g b^2\, ,\qquad \tom^1=-\b b^1\, ,       \nn\\
&&\tom^2=-\a b^0+\b b^2\, ,                                \nn
\eea
where $\a,\b,\g$ are defined in terms of the triad components as
follows:
\be
\a:=\frac{BN'}{N}\, ,\qquad \b:=\frac{BKC'}{2N}\, ,\qquad  \lab{A1}
\g:=\frac{BK'}{K}\, .
\ee
As a consequence, the Riemannian curvature $\tR^i$ takes the
following form:
\bea
&&\tR_0=\left(\b'B+2\b\g\right)b^0b^1
        -\left(\g'B+\g^2+\b^2\right)b^1b^2\, ,             \nn\\
&&\tR_1= -(\a\g+\b^2)b^2b^0\, ,                            \nn\\
&&\tR_2=-\left(\a'B+\a^2-3\b^2\right)b^0b^1
        -\left(\b'B+2\b\g\right)b^1b^2\, .                 \lab{A2}
\eea

Using the explicit form of $\tR_i$, we find that the essential
content of the second gravitational field equation \eq{3.7} is given
by the following four equations:
\bea
&&\b'B+2\b\g=\frac{1}{2}VEH\, ,                            \nn\\
&&\g'B+\g^2+\b^2=-\Leff+\frac{1}{2}VH^2 \, ,               \nn\\
&&\a'B+\a^2-3\b^2=-\Leff-\frac{1}{2}VH^2 \, ,              \nn\\
&&\a\g+\b^2=-\Leff\, .                                     \lab{A3}
\eea
In the Riemannian limit $u\to 0$, or equivalently $V\to -1/a=-2$ (in
units $8\pi G=1$), these equations reduce to equations (7)-(10) of
Ref. \cite{14}. Substituting the self-duality condition
$2\eps\b=\g-\a$ into the last equation, we find that $\g$ and $\a$
can be expressed in terms of $\b$ as
\bsubeq\lab{A4}
\be
\g=\frac{1}{\ell}+\eps\b\,,\qquad\a=\frac{1}{\ell}-\eps\b\,,\lab{A4a}
\ee
where we used $\Leff=:-1/\ell^2$. This implies, in particular,
\be
\frac{B(NK)'}{2N}=\frac{1}{2}(\a+\g)=\frac{1}{\ell}\, .    \lab{A4b}
\ee
\esubeq

Now, one can see that only one of the first three equations is
independent (we take it to be the first one). Using the radial
coordinate $\r$ introduced in Section 3 by $d\r=dr/B$, and the
relations \eq{3.5b} and \eq{3.4}, the first equation in \eq{A3}
reduces to the form
\be
\b'+2\b\frac{K'}{K}=
        \frac{\eps Q_e^2V}{2K^2}\exp(-2\eps\m\r)\, ,       \nn
\ee
where prime denotes now the derivative with respect to $\r$.
Substituting here the expression for $\b$ obtained from \eq{A4},
$\b=\eps(K'/K-1/\ell)$, we find
\be
(K^2)''-\frac{2}{\ell}(K^2)'=Q_e^2V\exp(-2\eps\m\r)\, .    \lab{A5}
\ee
This equation can be easily solved for $K^2$. Indeed,
\bsubeq\lab{A6}
\be
K^2=K^2_h+K^2_p\, ,
\ee
where $K^2_h$ is the general homogeneous solution, and
$K_p$ a particular solution of \eq{A5}:
\bea
&& K^2_h=g_1+g_2\exp(2\r/\ell)\, ,                         \nn\\[3pt]
&&\m=0:\quad K^2_p=-\frac{\ell VQ_e^2}{2}\r\, ,            \nn\\[3pt]
&&\m\ne 0:\quad
          K^2_p=\frac{VQ_e^2\ell^2/4}{\m^2\ell^2+\eps\m\ell}
                      \exp{(-2\eps\m\r)}\, .
\eea
\esubeq
For $\m\ne 0$, we also have to assume $\eps\m\ell\ne -1$.

Once we have $K^2$, we can go back to \eq{A4} and \eq{3.5b} and
calculate $N,C$ and $E,H$ (when the radial coordinate is $\r$, we
have $B=1$). Consequently:
\bitem
\item[(c)] The general solution of the system of field equations \eq{A3} and
\eq{3.3} for $\eps\m\ell\ne -1$, is determined by a single function
$K^2$.
\eitem
This is a remarkable dynamical feature of the $\m$-modified self-dual
Maxwell field in 3D gravity with torsion. It shows, in particular, a
complete resemblance with the corresponding Riemannian dynamics
\cite{14}, in accordance with the theorem (b) in Section 3.

\section{Calculation of the conserved charges \eq{4.3}}
\setcounter{equation}{0}

Our approach to the conserved charges is based on the canonical
formalism, as outlined in Section 4.  We start by giving the
asymptotics for the regularized field $b^i$,
\be
b^i{_\m}\sim\left( \ba{ccc}
  \dis\frac{r}\ell-\frac{r_0^2}{\ell r} & 0 & 0 \\
  0 & \dis\frac{\ell}{r}+\frac{\ell r_0^2}{r^3} & 0 \\
  \dis\frac{\eps r_0^2}{\ell r} & 0 & r
                      \ea
             \right)\, ,                                   \nn
\ee
for $\om^i$,
\bea
&&\tom^i{_\m}\sim\left( \ba{ccc}
  \dis-\frac{\eps r_0^2}{\ell^2r} & 0 & -\dis\frac{r}\ell \\
  0 & 0 & 0 \\
  -\dis\frac{r}{\ell^2}+\frac{r_0^2}{\ell^2 r} & 0 & 0
                        \ea
                        \right) \, ,                       \nn\\[6pt]
&& t^i{_\m}\sim \left( \ba{ccc}
  -\dis\frac{Q_e^2}{\ell r} & 0 & -\dis\frac{\eps Q_e^2}{r} \\
  0 & 0 & 0 \\
  \dis\frac{\eps Q_e^2}{\ell r} & 0 & \dis\frac{Q_e^2}{r}
                      \ea
                      \right)\, ,                           \nn\\[3pt]
&&\om^i{_\m}\sim\tom^i{_\m}+\frac{1}{2}(p b^i{_\m}+ut^i{_\m})\, ,\nn
\eea
and for $A$:
$$
A_\m\sim 0\, .
$$
Note that the logarithmic terms are hidden in higher-order terms. In
order to ensure consistency, we restrict the original gauge
parameters to the form compatible with the adopted asymptotic
conditions:
$$
\xi^\m=(\ell T_0,0,S_0)\, , \qquad
\th^i=(0,0,0)\, ,\qquad  \l=\l_0\, ,
$$
where $T_0,S_0$ and $\l_0$ are constant parameters associated to
the rigid time translations, axial rotations and $U(1)$ transformations,
respectively.

The calculation of the conserved charges is closely related to the
form of the canonical generator $G$. In the asymptotic region, where
the transformation parameters are constant, $G$ has the following
effective form:
\bea
&&G=-G_1-G_3 \, ,                                          \nn\\
&&G_1:=\xi^\r\left[b^i{}_\r\hat\cH_i+\om^i{}_\r\hat\cK_i
  +(\pd_\r b^i_0)\pi_i{}^0
  +(\pd_\r\om^i{}_0)\Pi^i{}_0+(\pd_\r A_0)\pi^0 \right]\,, \nn\\
&&G_3:=-\l\pd_\a\pi^\a\, ,                                 \nn
\eea
were $\hat\cH_i$ and $\hat\cK_i$ are components of the total
Hamiltonian (see Appendix C in Ref. \cite{8}).

The asymptotic generator $G$ acts on basic dynamical variables via
the Poisson brackets, and consequently, it must be differentiable. If
this is not the case, the form of $G$ can be improved by adding a
suitable surface term. To improve $G$, we calculate its variation:
\bea
&&\d G_1
  =\xi^\r\left\{-2\ve^{0\a\b}\pd_\a\left[b^i{_\r}\d(a\om_{i\b}
    +\a_4 b_{i\b})+\om^i{_\r}\d(ab_{i\b}+\a_3\om_{i\b})\right]
    +\d\t^0{_\r}\right\}+R\, ,                             \nn\\
&&\d G_3=-\l\pd_\b\d\pi^\b \, ,                            \nn
\eea
where $R$ denotes the contribution of regular (differentiable) terms.
Using the adopted asymptotic conditions, performing the integration
over the boundary $S_\as$ and taking the limit $r_\as\to\infty$, we
obtain:
\bea
&&\d G_1[\xi^0]=\xi^0\frac{2\pi\eps}{\ell}\d
  \left[\a_3 Q_e^2V(-\eps)\right]\, ,                      \nn\\
&&\d G_1[\xi^2]=\xi^2 2\pi\d\left[\a_3Q_e^2V(-\eps)\right]\,,\nn\\
&&\d G_3[\l]=2\pi\l\d Q_e\, .                              \nn
\eea
Thus, the improved canonical generator takes the form
$$
\tilde G=G+\xi^0 E+\xi^2 M+\l Q\, .
$$
where $E,M$ and $Q$ are the canonical charges of the self-dual
solution, displayed in \eq{4.3}.

\end{document}